\def\withFigures{1}  % if we want figures included, replace 0 -> 1
\def\figuresDir{}

\documentclass[english,amsmath,amssymb,floatfix,notitlepage,preprintnumbers,showpacs]{revtex4-1}

\usepackage{graphicx}

\makeatletter

\usepackage{babel}

\begin{document}
\preprint{Alberta Thy 13-11}

\title{Binding energy of the positronium negative ion via dimensional scaling}
\author{Nikita Blinov}
\altaffiliation[Current address: ]{Theory Group, TRIUMF, 4004 Wesbrook Mall, Vancouver, BC V6T 2A3,
Canada}
\affiliation{Department of Physics, University of Alberta, Edmonton, AB T6G 2E1, 
Canada}

\author{Andrzej Czarnecki}
\affiliation{Department of Physics, University of Alberta, Edmonton, AB T6G 2E1, 
Canada}
\begin{abstract}
  We determine  the binding energy of the negative positronium ion in
  the limits of one spatial dimension and of infinitely many
  dimensions. The numerical result for the one-dimensional ground
  state energy seems to be a rational number, suggesting the existence
  of an analytical solution for the wave function.  We construct a
  perturbation expansion around the infinitely-dimensional limit to
  compute an accurate estimate for the physical three-dimensional
  case.  That result for the energy agrees to five significant figures
  with variational studies.
\end{abstract}
\maketitle

\section{Introduction}
The negative positronium ion Ps$^{-}$ is a bound state of two
electrons and a positron. It is the simplest bound three body system
from the theoretical point of view, since it does not contain a hadronic
nucleus. It provides an important
testing ground for quantum electrodynamics (QED), which should be able
to describe this purely leptonic bound state with high precision.

Because of the $e^{+}e^{-}$ annihilation, Ps$^-$ is unstable, with a
lifetime of about four times that of para-positronium.  It decays
predominantly into two or three photons, with the one-photon decay
possible but extremely rare.   It is weakly bound and has no excited states in the discrete
spectrum \cite{PhysRevA.24.3242,PhysRevA.80.054502} (for a discussion of resonances, see \cite{Basu2011,Kar2011119}). 

Following its prediction
by Wheeler in 1946 \cite{wheeler1946} and experimental observation in
1981 by Mills \cite{mills1981}, the positronium ion has
been subject to much theoretical study. Its non-relativistic bound
state energy, decay rate, branching ratios of various
decay channels, and  polarizabilities
have been computed accurately using variational methods
\cite{bhatia1983,PhysRevA.48.4780,Frolov99,frolov2006,Frolov2007,drake,Puchalski:2007ck,PhysRevA.75.062510}. 

Recently, intense positronium sources have become available, opening
new possibilities for  experimental studies of Ps$^-$ \cite{canexp}.
The  measured decay rate  \cite{mills1983,fleischer2006} agrees
with the theoretical prediction.
Improved measurements of the decay rate, the three-photon branching
ratio, and the binding energy have been proposed \cite{FleischerInKarsh}.

A challenge in the theoretical study of this three-body system is that
its wave function is not known analytically, even if only the Coulomb
interaction is considered.  Since all particle masses and magnitudes
of their charges are equal, it is not possible to use the
Born-Oppenheimer approximation.  So far  all precise theoretical
predictions of Ps$^-$ properties have relied on variational calculations.

In the present paper we explore a different approach to computing the
wave function and the binding energy of Ps$^-$.  We use dimensional
scaling (DS) method, in which the dimensionality of
space $D$ is a variable.  We focus on the limits $D\rightarrow1$ and $D\rightarrow\infty$.
A precise result for $D=3$ may be obtained by interpolating between
the two limits using perturbation theory in $1/D$. The advantage
of DS is that the two limits of the Schr\"{o}dinger equation often
have relatively simple solutions.  Full inter-particle correlation effects are included at every
order in the perturbation expansion in $1/D$. 
More information about dimensional
scaling and further references can be found in
\cite{dimscale,dunn:5987,goodson:8481,PhysRevA.46.5428,PhysRevA.51.R5,witten1980}.

It is important to note that the dimensional limits considered here
are not physical in the sense that the form of the potential energy
is taken to be $1/r$, regardless of the dimension.  A physical limit
of a system would use an appropriate Coulomb potential that is the
solution of a $D$-dimensional Poisson equation. For example for $D=1$
it is linear, logarithmic for $D=2$, and depends on charge separation
as $r^{-(D-2)}$ for $D>2$. Since we are ultimately interested in
$D=3$ physics, it is useful to fix the potential to be the $D=3$
Coulomb interaction. The $D\rightarrow1$ limit used here offers the
additional simplification that after coordinate and energy rescaling
the potential takes form $(D-1)/r$ , which can be formally replaced
by a Dirac delta function  \cite{PhysRevA.34.2654}.

We find that the DS provides a useful complement to the variational
method.  In the future, it can be employed to independently check matrix elements of
operators needed in precise studies of Ps$^-$. 

This paper is organized as follows. In Section~\ref{sec:d1} we consider
the $D=1$ limit of the Ps$^{-}$ system.
We solve the Schr\"{o}dinger equation numerically to find an eigenvalue that approaches
a simple rational number, possibly hinting at the existence of an
analytical solution. 

In Section~\ref{sec:dinfty} we consider the $D\rightarrow\infty$
limit and describe the resulting $1/D$ expansion. 
We sum  up the perturbation
series for the ground state energy and evaluate it at $D=3$.  The
binding energy we find agrees with variational studies to five
significant figures. We conclude
in Section~\ref{sec:Conclusion}.

\section{$D=1$ Limit of $\mbox{Ps}\mbox{\ensuremath{^{-}}}$\label{sec:d1}}
In the one dimensional limit, the Coulomb
potential is represented by the Dirac delta function \cite{PhysRevA.34.2654}.
Delta function models have been used extensively also in condensed
matter physics. A simple analytical wave function exists for
any number of identical particles interacting via attractive potentials
\cite{McGuire1964}. The case of all repulsive potentials with periodic
boundary conditions has been treated by Lieb and Liniger \cite{Lieb1963},
and Yang \cite{Yang1967}. More recent works have studied one dimensional
systems with both attractive and repulsive delta interactions. Craig
\emph{et al.} considered the dependence of the energy on the number
of particles in a system of equal numbers of positively and negatively
charged bosons \cite{Craig1992}. Li and Ma studied a system of $N$
identical particles with an impurity with periodic boundary conditions
\cite{Li1995}. 

The $D=1$ limit of the Ps$^{-}$ quantum problem
is a delta function model with two attractive and one repulsive delta
functions with non-periodic boundary conditions, which, to the best
of our knowledge, has not yet been solved. We present a derivation
of a one dimensional integral equation for the solution to this problem,
analogous to the helium case treated by Rosenthal \cite{rosenthal}.

The time independent Schr\"{o}dinger equation for the relative motion
of $\mbox{Ps}^{-}$ takes the dimensionless form \begin{equation}
\left(-\frac{1}{2}\left[\nabla_1^{2}+\nabla_2^{2}+\nabla_1\cdot\nabla_2\right]
-\frac{1}{r_1}-\frac{1}{r_2}+\frac{1}{r}\right)\psi
=\varepsilon\psi,
\label{eq:SchrodEqExplicit}
\end{equation}
where $r_1$ and $r_2$ are the electron-positron distances, $r$
is the inter-electron distance (in units of $2(m\alpha)^{-1}$ with
$\hbar=c=1$) and $\varepsilon$ determines the energy eigenvalue, 
$E =\varepsilon m\alpha^{2}/2$.  This choice of units helps compare 
intermediate results with Rosenthal's delta function model
of helium \cite{rosenthal}. 

In the limit $D\rightarrow1$, we let $\vec{r}_1\rightarrow x$ and
$\vec{r}_2\rightarrow y$, where $-\infty<x,\, y<\infty$; the
gradients become partial derivatives and the Coulomb potentials are
replaced by Dirac delta functions (this limit is described in detail
in \cite{twoelec1d}).
Equation (\ref{eq:SchrodEqExplicit}) is replaced by
\begin{eqnarray}
&& \hspace*{-1mm}
\left[-\frac{1}{2}
\left( 
\frac{\partial^{2}}{\partial    x^{2}}
+\frac{\partial^{2}}{\partial    y^{2}}
+\frac{\partial^{2}}{\partial x\partial    y}\right)
-\delta(x)-\delta(y)+\delta(x-y)
\right]\psi 
 \nonumber \\
&&  \hspace*{10mm}= \varepsilon\psi.
\label{eq:PosShrodEq}\end{eqnarray}
Using Fourier transformation, we rewrite this Schr\"odinger equation
as a one-dimensional integral equation,
\begin{equation}
G(k_1,k_2)=\frac{F(k_1)+F(k_2)-H(k_1+k_2)}{\frac{1}{2}(k_1^{2}+k_2^{2}+k_1k_2+p^{2})},\label{eq:Gkk}\end{equation}
where the Fourier transforms of the wave function
$\psi(x,y)$ are %
\begin{eqnarray}
G(k_{1,}k_2) &=&\int\!\!\!\int e^{-ik_1x-ik_2y}\psi(x,y)dxdy,\label{eq:transG}\\
 F(k) &=& \int e^{-ikx}\psi(x,0)dx,\label{eq:xPsiTran} \\
H(k) &=& \int e^{-ikx}\psi(x,x)dx,\label{eq:xyPsiTran}
\end{eqnarray}
and $p^{2}/2=-\varepsilon$. 
We now invert the transformation (\ref{eq:transG}), and use the
resulting $\psi$ in eqs.~(\ref{eq:xPsiTran},\ref{eq:xyPsiTran}) to
obtain a system of two integral equations for $F(k)$
and $H(k)$. These are easily decoupled and yield 
\begin{widetext}\begin{eqnarray}
F(k) & = & \frac{2F(k)}{\sqrt{3k^{2}+4p^2}}
+\frac{1}{\pi}\int\frac{F(k') dk'}{k^{2}+k'^{2}+kk'+p^{2}} \nonumber \\
 & - &
 \frac{2}{\pi^{2}}\int
\frac{\sqrt{3k'^{2}+4p^2}}{2+ \sqrt{3k'^{2}+4p^2} }
\frac{1}{k^{2}+k'^{2}-kk'+p^{2}}\left(\int\frac{F(k'') dk''}{k'^{2}+k''^{2}-k'k''+p^{2}}\right)dk',
\label{eq:PsIntegralEquationExplicit}
\end{eqnarray}
\end{widetext}and\begin{equation}
H(k)=\frac{2}{\pi}
\frac{\sqrt{3k^{2}+4p^2}}{2+ \sqrt{3k^{2}+4p^2} }
\int\frac{F(k') dk'}{k^{2}+k'^{2}-kk'+p^{2}}.\label{eq:Hk-1}\end{equation}
Once $F(k)$ is found one can compute $H(k)$. The  two-dimensional
eigenvalue problem is thus reduced to a one-dimensional integral equation
(\ref{eq:PsIntegralEquationExplicit}), which we solve numerically.
The integral equation is discretized using Gauss-Legendre
quadrature, casting it into a system of homogeneous linear equations
for $F(k_i)$, where $k_i$ are the abscissas. The system has
a non-trivial solution when the determinant of the discretized integral
kernel vanishes. This condition fixes the value of $p$ and thus the
$D=1$ binding energy. 

The wave function is then determined by solving
the linear system for $F(k_i)$. One finds that $F(k_i)$ spans
the null space of the discretized kernel and can be computed using
its singular value decomposition.
We used cubic spline interpolation on the set $\{F(k_i)\}$
to interpolate between the quadrature points 
and generate an approximation for $F(k)$. 

Once $F(k)$ is known, the functions $H(k)$ and $G(k_1,k_2)$
are constructed using eq.~(\ref{eq:Hk-1}) and (\ref{eq:Gkk}). 
Finally, the wave function $\psi(x,y)$ is  obtained by the inverse
Fourier transformation of $G(k_1,k_2)$.

We performed this procedure for various quadrature sizes $N$ with
the results summarized in Table \ref{tab:The-bound-state}. %

\begin{table}[htb]
\begin{ruledtabular}
\begin{tabular}{cc}
Quadrature size $N$ & $\varepsilon$\tabularnewline
$10$ & $-0.6666657902370426$\tabularnewline
$20$ & $-0.6666666661283767$\tabularnewline
$50$ & $-0.6666666666666257$\tabularnewline
$100$ & $-0.6666666666666660$\tabularnewline
\end{tabular}
\end{ruledtabular}
\caption{Binding energy of the one-dimensional model of the
  positronium ion, in
units $m\alpha^{2}/2$. For $N\geq100$ the eigenvalue appears to converge
to $-2/3$. For these large quadrature sizes the uncertainty in the energy
is in the last digit due to finite precision used in the calculation.
\label{tab:The-bound-state}}
\end{table}

\begin{figure*}
\begin{centering}
%\subfloat[]{\begin{centering}
\if\withFigures1
\includegraphics{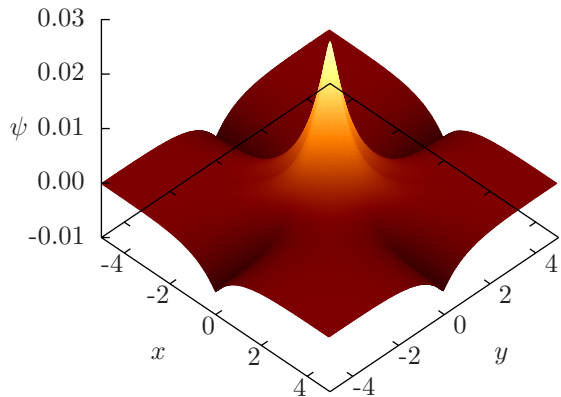}
\fi
%\par\end{centering}

%}\subfloat[]{\begin{centering}
%\if\withFigures1
%\includegraphics{\figuresDir wf3Dcontour}
%\fi
%\par\end{centering}

%}
\par\end{centering}

\caption{(Color online) The unnormalized wave function $\psi(x,y)$ satisfying eq.~(\ref{eq:PosShrodEq}).  The distances are in units $2/(m\alpha)$.
\label{fig:wf3D}}

\end{figure*}

\begin{figure*}
\begin{centering}
%\subfloat[]{\begin{centering}
\if\withFigures1
\includegraphics{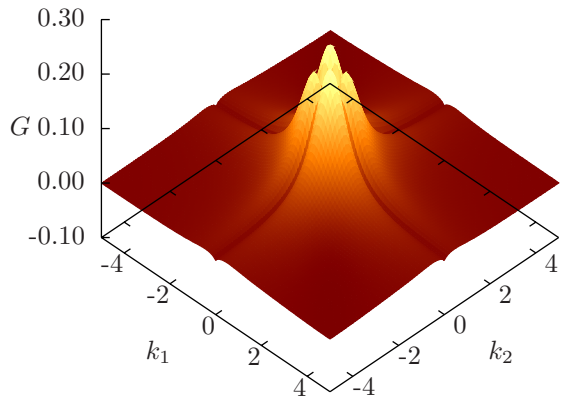}
\fi
%\par\end{centering}

%}\subfloat[]{\begin{centering}
%\if\withFigures1
%\includegraphics{\figuresDir wfFT3Dcontour}
%\fi
%\par\end{centering}

%}
\par\end{centering}

\caption{(Color online) The Fourier transform $G(k_1,k_2)$ of the ground state position-space
wave function $\psi(x,y)$ of eq.~(\ref{eq:PosShrodEq}) computed numerically.
The Fourier space coordinates have units $m\alpha /2$.
\label{fig:wfFT}}

\end{figure*}

We see that as $N$ increases, $\varepsilon$ approaches $-2/3$. For $N=100$
the $16$ decimal place precision limit of the \verb=double= data
type used in the calculation is almost reached. This simple numerical
result
suggests that the one dimensional  Schr\"{o}dinger equation has
an analytical solution.
The wave function and its Fourier transform are plotted in Figures
\ref{fig:wf3D} and \ref{fig:wfFT}. We observe that the wavefunction
has ridges at $x=0$, $y=0$ and $x=y$ as expected from the delta
function potential of eq.~(\ref{eq:PosShrodEq}). A simple numerical
comparison of $H(k)$ with the Fourier transform of $\exp(-a|x|)$
indicates that the wavefunction fall-off in the $x=y$ direction
is nearly exponential.%

%
% Fig 1

% Fig 2

The result $\varepsilon=-2/3$ translates into the energy eigenvalue $E$ equal
$-1/3$ atomic unit of energy (1 a.u. $=m\alpha^2$) or $-9.07$
eV.  This is in qualitative agreement with the actual value that is
about $-0.26$ a.u.,
just below $-1/4$ a.u.~(this  fraction is the binding energy of a
positronium atom in the non-relativistic approximation).

We note that the eigenvalue that can be obtained for the two-body
problem (the positronium or the hydrogen atom) in the one dimensional
delta model coincides precisely with the physical value.  This is the
case because the wave function in the delta model has the same cusp at
the origin as the radial wave function in the physical space.  Thus
the delta model reproduces that radial wave function exactly.  For the
three-body problem the agreement is only rough.

It would be interesting to determine the one dimensional wave function
analytically.  We remark that the Schr\"odinger equation of the three
body problem
\eqref{eq:PosShrodEq} can be rewritten, with a simple change of
variables $x,y$, in the form of a one-particle motion in the external
potential consisting of two attractive and one repulsive delta
function ridges.

In the following section we focus on the opposite limit of very many
dimensions.  We shall find that an expansion around that limit can be
constructed, giving a very accurate determination of the binding
energy of $\mathrm{Ps}^-$.  Interestingly, the $D=1$ method will be
again useful: it will provide an important subtraction term that we
will use to accelerate the convergence of a perturbative expansion.

\section{$D\rightarrow\infty$ and Dimensional Perturbation Theory\label{sec:dinfty}}

The first step in taking the $D\rightarrow\infty$ limit is to generalize
the $\mbox{Ps\ensuremath{{}^{-}}}$ Schr\"{o}dinger equation to $D$
dimensions. We are interested in the ground state, which is completely
described by the three inter-particle distances $\rho_{ij}=|\vec{r}_i-\vec{r}_{j}|$.
The Schr\"{o}dinger equation takes the form
\begin{equation}
H\phi\equiv(T+U+V)\phi=E\phi,
\end{equation}
where $E$ is the energy in atomic units $m\alpha^{2}$  
(note that it differs by a factor 1/2 from the $\varepsilon$
used in  $D=1$ in the previous section) and 
\begin{eqnarray}
T&=&-\frac{1}{2}\sum_{i\neq j}\left(\frac{\partial^{2}}{\partial
    \rho_{ij}^{2}}+\sum_{k\neq
    i,j}\frac{\rho_{ij}^{2}+\rho_{ik}^{2}-\rho_{jk}^{2}}{2\rho_{ij}\rho_{ik}}\frac{\partial^{2}}{\partial
    \rho_{ij}\partial \rho_{ik}}\right),
\nonumber
\\
U&=&\frac{(D-1)(D-5)}{8\Upsilon^{2}}\left(\rho_{13}^{2}+\rho_{23}^{2}+\rho_{12}^{2}\right),
\nonumber
\\
V&=&\frac{1}{\rho_{12}}-\frac{1}{\rho_{13}}-\frac{1}{\rho_{23}},\label{eq:pertPot}
\end{eqnarray}
 and $\phi=\Upsilon^{(D-1)/2}\psi$ is the rescaled wave function with
\begin{eqnarray}
\Upsilon &=& 2\sqrt{s(s-\rho_{12})(s-\rho_{13})(s-\rho_{23})},\label{eq:area}
\\
s&=&\frac{1}{2}(\rho_{12}+\rho_{13}+\rho_{23}).\label{eq:ess}
\end{eqnarray}
Note that the characteristic $D^{2}$ dimensional dependence is confined
to $U$. (The $U$ term in the effective potential is the usual centrifugal
contribution from the kinetic energy found by expressing the Laplacian
in terms of $\rho_{ij}$.) In order to obtain a finite limit the coordinates and the
energy must be rescaled, $\rho_{ij} = D^{2}r_{ij}$ and
$E=\epsilon/D^2$. This introduces a factor of
$1/D^{2}$ in front of the kinetic energy term, eq.~(\ref{eq:pertPot}),
so in the limit $D\rightarrow\infty$ it is suppressed. In terms of
the rescaled quantities, the Schr\"{o}dinger equation is written
as 
\begin{equation} 
(\delta^{2}T+\delta^{2}U+V)\phi=\epsilon\phi,\label{eq:psminscaledSchrod}\end{equation}
where $\delta=1/D.$ %

In the limit $\delta\rightarrow0$, terms containing derivatives
vanish in eq.~(\ref{eq:psminscaledSchrod}). Since the ground state
energy is the smallest eigenvalue of the Hamiltonian, we seek to minimize
the effective potential 
\begin{equation}
V_\mathrm{eff}=\delta^{2}U+V\label{eq:effectivepot}\end{equation}
at $\delta=0$, under the constraint that $r_{ij}$ define a triangle.
Unfortunately, in $D\to \infty$, the Ps$^-$ system described by the
potential in eq.~(\ref{eq:pertPot}) is unbound (even if the positron were very
heavy, its charge would have to be larger than 1.228 for a bound state
to exists \cite{PhysRevA.51.R5} (see also \cite{Hogaasen:2010,PhysRevA.71.052505})).  The qualitative explanation of this
is that even though we have increased the number of spatial
dimensions, we have retained the $1/r$ behavior of the Coulomb
potential.  Thus it is relatively stronger at large distances than in
three dimensions and the electron-electron repulsion plays a more
important role even if the electrons are on the opposite sides of the positron.

However, the strict $\delta=0$
regime is unphysical. We are interested in the $\delta=1/3$ case,
so we are free to modify the potential as long as it reduces
to the correct form at $\delta=1/3$. This can be done by reducing
the strength of the electron-electron repulsion, as was done for H$^{-}$
\cite{PhysRevA.51.R5},
\begin{equation}
V=\frac{\lambda_{0}+3(1-\lambda_{0})\delta}{r_{12}}-\frac{1}{r_{13}}-\frac{1}{r_{23}},
\label{eq:modpot}\end{equation}
where $\lambda_{0}$ is a free numerical parameter. Note that at
$\delta=1/3$, 
eq.~(\ref{eq:modpot}) reduces to eq.~(\ref{eq:pertPot}), as required.
We have used $\lambda_{0}=0.5$ throughout our computations since
this value gave the best results for the H$^{-}$ system, but other
values of $\lambda_{0}$ may result in better convergence 
of the perturbation series for $\mbox{Ps}^{-}$. The effective potential,
eq.~(\ref{eq:effectivepot}), is minimized at 
\begin{eqnarray}
\bar{r}_{12} &=&0.773603828324(1),\nonumber \\
%03828324456$,
\bar{r}_{13}=\bar{r}_{23}&=&0.5866862922582(4),
% 0.586686292258617$ 
  \label{eq:rMin}
\end{eqnarray}
 with the minimum value of 
$V_{0}=-1.381325607963162(1)$.
%$V_{0}=-1.381325607963162$.
The errors are estimated by performing the calculation again with  higher precision and
smaller tolerances. Convergence is
ensured by restarting the minimization from a slightly perturbed location. 
We note that this result corresponds to the energies rescaled by
$\delta^2$.  Thus, to compare with the physical value, we have to
divide this result by $3^2=9$, obtaining the first estimate of the
binding energy $\simeq -0.15$ a.u., to be compared with the known
value (see Table \ref{tab:Results-of-summation}) of
about $-0.26$ a.u.

The static $\delta=0$ limit is the zeroth order in $1/D$ expansion
but, without the kinetic energy, it does not allow us to generate
further orders in the perturbation expansion. In order to construct
such an expansion, we consider the next simplest case, the harmonic
approximation to the potential. This will yield a complete set of
states that can be used to generate an expansion. The natural
expansion parameter for eq.~(\ref{eq:psminscaledSchrod}) is
$\delta^{1/2}$.  This follows from the dominant balance argument
applied to the Schr\"{o}dinger equation. One finds that, for $\delta^{1/2}$, the
harmonic terms in the expansion of the potential are of the same
order as the constant coefficient terms in the kinetic energy
expansion.

Details of the procedure used to construct the expansion are described
in the Appendix.
The summation of the resulting series in powers of $\delta=1/D$ is complicated by the
fact that the expansion is divergent at high orders due to a singularity
at $\delta=0$ \cite{elout:5112}, so we expect the convergence of
the naive summation \begin{equation}
E(\delta)=\delta^{2}\sum_{k=0}^{\infty}E_{k}\delta^{k}\end{equation}
to be slow. In the above expression $E_{0}=V_{0}$ and
$E_k=\epsilon_{2k-2}$ for $k>0$, where $\epsilon_k$ are 
expansion coefficients of the rescaled energy that appears in eq.~(\ref{eq:psminscaledSchrod}).
There are also poles at $D=1$ that slow down the asymptotic
convergence of the expansion at low values of $D$. A better
estimate for $E$ can be obtained by subtracting these poles from the
expansion. To this end, the residues of the poles must be determined.
Following
\cite{PhysRevA.51.R5,elout:5112} we define 
\begin{equation}
E(\delta)=\delta^2
\left[\frac{a_{-2}}{(1-\delta)^2}
+\frac{a_{-1}}{1-\delta}
+\sum_{k=0}^\infty E_{k}'\delta^k\right],
\label{eq:energyExpansion3}\end{equation}
 where 
\begin{equation}
E_{k}'=E_k - (k+1)a_{-2} - a_{-1}.
\end{equation}
The residue of the second order pole, $a_{-2}$, corresponds to the ground state energy in the $D=1$
limit (more precisely $a_{-2}=4E_{D=1}$).
We have computed it employing again the method described in Section
\ref{sec:d1}, this time with the rescaled charges of electrons and the
positron so as to satisfy eq.~(\ref{eq:modpot}). We find
\begin{equation}
a_{-2}=-1.102499999999999(1),
\label{eq:secondorderPole}\end{equation}
which again (see Table \ref{tab:The-bound-state}) 
resembles a rational number, indicating that there are
likely analytical solutions of the $D=1$ model even for an arbitrary 
charge of the positron (not necessarily equal in magnitude to that of
the electron). As in Table \ref{tab:The-bound-state}, the uncertainty
in this converged residue is due to finite precision, as was checked
by using larger quadrature sizes.

To find the residue of the single pole, $a_{-1}$, we 
subtract the double pole from both sides of
eq.~(\ref{eq:energyExpansion3}) and multiply by $1-\delta$. We get
the condition
\begin{equation}
a_{-1}=\lim_{\delta\rightarrow1}\sum_{k=0}^{\infty}(E_k - E_{k-1}-a_{-2})\delta^k,
\label{eq:firstorderPole}\end{equation}
where $E_{-1}=0$. In practice we only have a finite number of
terms in the sums in equations (\ref{eq:energyExpansion3}) and
(\ref{eq:firstorderPole}). 
Pad\'e approximants have been shown to work well for summing up $1/D$
expansions \cite{PhysRevA.46.5428,PhysRevA.51.R5}. Using this method
to compute the limit in eq.~(\ref{eq:firstorderPole}) we get for
$\mbox{Ps}\mbox{\ensuremath{^{-}}}$,
\begin{equation}
a_{-1}=0.427(2).
\label{eq:a1res}\end{equation}
This result was obtained with the first 21 terms in the sum
in eq.~(\ref{eq:firstorderPole}). The uncertainty in the computed 
value was estimated by varying the order of the Pad\'e approximant for $a_{-1}$ 
as $[N/M]\rightarrow[(N-1)/(M+1)]$  \cite{PadeEncyclopedia}. If the result has converged, the order
of the approximant should not matter (barring the introduction of
spurious poles in the denominator of the approximant). We 
use this method to estimate the error for all quantities computed
using Pad\'e approximants.
In our calculations 
we use the full unrounded result for $a_{-1}$
which gives a slightly worse result for the bound state energy than
eq.~(\ref{eq:a1res}). 
As has been noted in Ref.~\cite{PhysRevA.46.5428}, 
this way of determining $a_{-1}$ is not very accurate. An exact value for $a_{-1}$ (in principle
obtainable from expansions about $D=1$) would improve the convergence
of the $1/D$ expansion. 
For He we used $a_{-2}=-3.15546$~\cite{rosenthal} to get $a_{-1}=0.313(1)$
using an identical calculation
with 21 energy expansion coefficients. We then evaluated eq.~(\ref{eq:energyExpansion3})
(with the summation truncated again at 21 terms) at $\delta=1/3$.
For helium, this yields a ground state energy that agrees with the
variational calculation of \cite{korobov2002}
to five digits, which is consistent with the result of \cite{goodson:8481}
for this summation method and perturbation expansion cutoff. The same
calculation for the positronium ion yields a five digit agreement with
the results in \cite{frolov2006,Frolov2007,Puchalski:2007ck}. These results
are summarized in Table \ref{tab:Results-of-summation}. 
\begin{table}[htb]
\begin{ruledtabular}
\begin{tabular}{ccc}
%\hline 
 & Known energy (1 a. u. = $m\alpha^{2}$) & $1/D$ Expansion \tabularnewline
%\hline
%\hline 
He & $-2.9037243770341196$ \cite{korobov2002} & $-2.90374(1)$\tabularnewline
%\hline 
Ps$^{-}$ & $-0.2620050702329801$ \cite{Frolov2007} & $-0.262005(2)$\tabularnewline
%\hline
\end{tabular}
\end{ruledtabular}
\caption{Results of summation of the $1/D$ expansion using Pad\'e summation
with first and second order poles removed. The first $21$ non-zero
terms were used in the summation. \label{tab:Results-of-summation}}
\end{table}

\begin{figure}[htb]
\begin{centering}
\if\withFigures1
\includegraphics{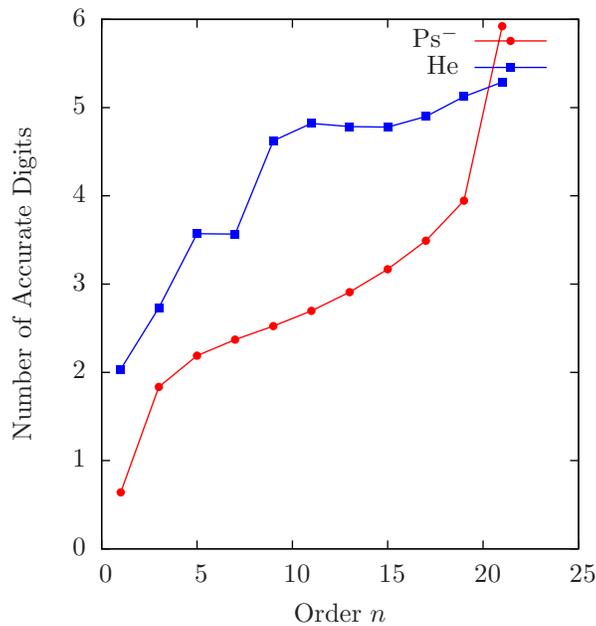}
\fi
\par\end{centering}

\caption{(Color online) Number of accurate digits in the ground state energy, defined
 as $-\log_{10}[(E-E_{\mbox{\footnotesize exact}})/E_{\mbox{\footnotesize exact}}]$
as a function of the number of terms in the summation of eq.~(\ref{eq:energyExpansion3}).\label{fig:convergence}}

\end{figure}

 Figure \ref{fig:convergence} shows the improvements to the energy
that are obtained by summing more terms. Higher orders yield better
accuracy despite the poor behaviour of the $1/D$ expansion coefficients
(see Table \ref{tab:Energy-expansion-coeffients-latest}). In fact,
the pole subtraction and Pad\'e resummation described above are 
necessary to get a sensible answer.

Aside from computing
higher orders in perturbation theory, 
precision of the result may be
improved by using a different summation method. 
For example, Ref.~\cite{goodson:8481} found that Pad\'e-Borel
summation gives better results for helium than Pad\'e summation.

\section{Conclusions\label{sec:Conclusion}}

We have investigated the viability of dimensional scaling for making
accurate predictions for the positronium ion system. Equal masses
and correlation strengths make $\mbox{Ps}{}^{-}$ a good candidate
for the dimensional scaling treatment. We considered the $D=1$ limit and found that
the Schr\"{o}dinger equation can be reduced to a one dimensional
integral equation. The numerical solution for the energy eigenvalue
approaches a simple rational number suggesting the possibility of
a completely analytical solution. While this energy is not physically
relevant by itself, it can be used to accelerate the convergence of the
$1/D$ perturbation series. 

We constructed such a perturbative series by expanding
the solution of the full Schr\"{o}dinger equation about the $D\rightarrow\infty$
limit. Each coefficient was computed exactly in the harmonic basis.
To obtain an accuracy of five significant figures required expanding
up to order $41$ in perturbation theory. While the accuracy of the
energy expansion at this order is not yet competitive with variational
calculations, the present method provides a valuable alternative
approach to few body systems. It can be used to check a variety of
matrix elements that have previously been computed only
variationally.

In the future, higher orders in the $1/D$ perturbation series
can be determined without sacrificing speed if the analytical expansions
can be replaced with numerical evaluations of series coefficients
through finite differencing.
It would also be very valuable to establish how the convergence of this
expansion depends on the value of the parameter $\lambda_0$ introduced
in eq.~(\ref{eq:modpot}). Finally, the accuracy of the obtained wave
function should be determined by evaluating matrix element of various
operators and comparing them with the variational approach.

\begin{acknowledgments}

We thank Juan Maldacena for a discussion that initiated this study.
This research was supported by Science and Engineering Research Canada
(NSERC). 

\end{acknowledgments}

\appendix
\section{Perturbative expansion in $1/D$}
In this Appendix we describe how the coefficients of the $1/D$
expansion were determined.
Our procedure follows the matrix method of ref.~\cite{dunn:5987}.
In terms of the displacement coordinates $x_i$ defined by \begin{eqnarray}
r_{12} & = & \bar{r}_{12}+\delta^{1/2}x_1\nonumber \\
r_{13} & = & \bar{r}_{13}+\delta^{1/2}x_2\nonumber \\
r_{23} & = & \bar{r}_{23}+\delta^{1/2}x_3,\label{eq:transform}
\end{eqnarray}
($\bar{r}_{ij}$ are the coordinates of the minimum of the effective
potential, eq.~(\ref{eq:rMin})),
the Schr\"{o}dinger equation takes the form 
\begin{equation}
(\delta T+V_\mathrm{eff})\phi=\epsilon\phi.\label{eq:rdy2Expand}
\end{equation}
 The Hamiltonian is expanded in powers of $\delta^{1/2}$ such that
\begin{eqnarray}
T&=&\sum_{i=0}^{\infty}T_i\delta^{i/2},\label{eq:kineticExpansion}
\\
V_\mathrm{eff} &=&\sum_{i=0}^{\infty}V_i\delta^{i/2}.
\label{eq:effectivepotExpansion}\end{eqnarray}
Since the expansion is about the minimum of $V_\mathrm{eff}$, there is
no linear term in its expansion and $V_1=0$.
Also the kinetic energy $T$ starts contributing only in the second order in $\delta^{1/2}$,
so the energy $\epsilon$ has the form
\begin{equation}
\epsilon=V_{0}+\delta\sum_{i=0}^{\infty}\epsilon_i\delta^{i/2},
\label{eq:energyExpansion}
\end{equation}
For second order in $\delta^{1/2}$ in eq.~(\ref{eq:rdy2Expand}) we
have
\begin{eqnarray}
T_{0}&=&-{1\over 2}\sum_{i+j+k=2}t_{ijk}\left(\frac{\partial}{\partial
    x_1}\right)^i\left(\frac{\partial}{\partial
    x_2}\right)^{j}\left(\frac{\partial}{\partial
    x_3}\right)^{k}, \nonumber
%\label{eq:kinHarmonic}
\\
V_2&=&v_{000}+\sum_{i+j+k=2}v_{ijk}x_1^ix_2^{j}x_3^{k},\label{eq:potHarmonic}
\end{eqnarray}
where $t_{ijk}$ and $v_{ijk}$ are expansion coefficients that are
functions of $\bar{r}_{ij}$. %
This order in perturbation theory corresponds to three coupled harmonic
oscillators. To solve the Schr\"{o}dinger equation they need to be
decoupled.%
This procedure yields normal mode frequencies $\omega_i$ and the
corresponding normal coordinates $q_i$, related to $x_i$ by
a linear transformation $S$,
\begin{equation}
q_i=\sum_{j=1}^{3}(S^{-1})_{ij}x_{j}.\label{eq:normcoord}
\end{equation}
In terms of coordinates $q_i$, 
\begin{eqnarray}
T_{0} &=&-\frac{1}{2}\sum_{i=1}^{3}\frac{\partial^{2}}{\partial
  q_i^{2}},\label{eq:kinDiagonal-1}
\\
V_2&=&v_{000}+\frac{1}{2}\sum_{i=1}^{3}\omega_i^{2}q_i^{2}.\label{eq:potDiagonal}\end{eqnarray}
Defining \begin{equation}
H_i=T_i+V_{i+2},\label{eq:hamiltonianCoef}\end{equation}
the Hamiltonian can be written as \begin{equation}
H=\delta T+V_\mathrm{eff}=V_{0}+\delta\sum_{i=0}^{\infty}H_i\delta^{i/2}.\label{eq:hamiltonianExpansion}\end{equation}
Next we consider the wave function expansion \begin{equation}
\phi=\sum_{i=0}^{\infty}\phi_i\delta^{i/2}.\label{eq:wfExpansion}\end{equation}
Without loss of generality $\phi_i$ can be normalized as \begin{equation}
\langle\phi_{0}|\phi_{j}\rangle=\delta_{0,j}.\label{eq:wfCoefNormalization}\end{equation}
Collecting like powers of $\delta^{\frac{1}{2}}$ in eq.~(\ref{eq:rdy2Expand})
yields\begin{equation}
\sum_{i=0}^{p}(H_i-\epsilon_i)\phi_{p-i}=0.\label{eq:perturbationOrderP}\end{equation}
The $p=0$ order equation is a system of three independent harmonic
oscillators with the solution
\begin{eqnarray}
\phi_{0} &=&
h_{\nu_1}(q_1)h_{\nu_2}(q_2)h_{\nu_3}(q_3),\label{eq:harmonicWF}
\\
\epsilon_{0} &=& v_{000}+\sum_{i=1}^{3}\left(\nu_i+\frac{1}{2}\right)\omega_i,\label{eq:harmonicEnergy}\end{eqnarray}
with
\begin{equation}
h_{\nu}(q_i)=
%(\frac{\omega_i}{\pi})^{\frac{1}{4}}
\sqrt[4]{\frac{\omega_i}{\pi}}
\frac{1}{\sqrt{2^{\nu}\nu!}}H_{\nu}\left(\sqrt{\omega_i}q_i\right)e^{-\omega_iq_i^{2}/2},\label{eq:hermiteFunctions}\end{equation}
where $H_{\nu}$ is the $\nu$'th Hermite polynomial. For the ground state,
$\nu_i=0$. 

To compute further orders in the perturbation expansion,
$\phi_{j}$ from eq.~(\ref{eq:wfExpansion}) are projected onto the
harmonic oscillator basis
\begin{equation}
\phi_{j}=\sum_{i_1,i_2,i_3}{_{j}}a^{i_1i_2i_3}h_{i_1}(q_1)h_{i_2}(q_2)h_{i_3}(q_3).
\label{eq:harmonicProjection}\end{equation}
Here ${_{j}}a^{i_1i_2i_3}$ are the expansion coefficients.
The advantage of using the Hermite function basis is that only a finite
basis at every order of perturbation theory is needed, since the perturbations
are polynomials in $q_i$. Thus the perturbation expansion coefficients
can be computed exactly. We note that
$_{0}a^{i_1i_2i_3}=\delta_{0,i_1}\delta_{0,i_2}\delta_{0,i_3}$.
Equation (\ref{eq:wfCoefNormalization}) then implies that for any $p>0$\begin{equation}
{_{p}}a^{000}=0.\label{eq:nogstateOverlap}\end{equation}
The matrix elements of the operators $H_{j}$ defined in eq.~(\ref{eq:hamiltonianCoef})
are computed by noting that each $H_{j}$ is a sum of terms of the
form\begin{equation}
q_1^{i_1}q_2^{i_2}q_3^{i_3}\left(\frac{\partial}{\partial q_1}\right)^{\alpha_1}\left(\frac{\partial}{\partial q_2}\right)^{\alpha_2}\left(\frac{\partial}{\partial q_3}\right)^{\alpha_3},\end{equation}
where $\alpha_1+\alpha_2+\alpha_3=2$ for the kinetic terms
and $\alpha_i=0$ for terms coming from $V_\mathrm{eff}$. The matrix elements
of $q_i$ and $\frac{\partial}{\partial q_i}$ are derived from
the recurrence relations of the Hermite functions \cite{dunn:5987},
\begin{eqnarray}
q_i &=&\frac{1}{\sqrt{2\omega_i}}\left(\begin{array}{ccccc}
0 & \sqrt{1} & 0 & 0\\
\sqrt{1} & 0 & \sqrt{2} & 0\\
0 & \sqrt{2} & 0 & \sqrt{3} & \cdots\\
0 & 0 & \sqrt{3} & 0\\
 &  & \vdots &  & \ddots\end{array}\right),\label{eq:qMatrix}
\\
\frac{\partial}{\partial q_i} &=& \sqrt{\frac{\omega_i}{2}}\left(\begin{array}{ccccc}
0 & \sqrt{1} & 0 & 0\\
-\sqrt{1} & 0 & \sqrt{2} & 0\\
0 & -\sqrt{2} & 0 & \sqrt{3} & \cdots\\
0 & 0 & -\sqrt{3} & 0\\
 &  & \vdots &  & \ddots\end{array}\right),\label{eq:dqMatrix}\end{eqnarray}
 so $H_{j}$ is a linear combination of direct products of such matrices.
We denote the matrix representation of $H_{j}$ by $\mathbf{H}_{j}$,
and by $\mathbf{a}_{j}$ the tensor with elements $_{j}a^{i_1i_2i_3}$
in the harmonic basis. %
Finally we derive the recursion relations for computation of the energy
and wave function expansion coefficients. First, we rewrite eq.~(\ref{eq:perturbationOrderP})
in the harmonic basis \begin{equation}
\sum_{i=0}^{p}(\mathbf{H}_i-\epsilon_i)\mathbf{a}_{p-i}=0,\label{eq:perturbOrdPTensor}\end{equation}
and then contract with $\mathbf{a}_{0}$ and solve for $\epsilon_{p}$,
which yields\begin{equation}
\epsilon_{p}=\mathbf{a}_{0}\sum_{i=1}^{p}\mathbf{H}_i\mathbf{a}_{p-i}.\label{eq:energyRecursiveRel}\end{equation}
To compute the wave function expansion coefficients we need the pseudo-inverse
$\mathbf{K}$ of the operator $\mathbf{H}_{0}-\epsilon_{0}$, defined
component-wise as\begin{equation}
\mathbf{K}_{i_1i_2i_3}^{k_1k_2k_3}=\begin{cases}
0 & i_{\alpha}=k_{\alpha}=0\mbox{ }\forall\alpha\\
\left(\sum_{j=1}^{3}\omega_{j}i_{j}\right)^{-1}\delta_{i_1i_2i_3}^{k_1k_2k_3} & \mbox{otherwise}\end{cases}.\label{eq:pseudoInverse}\end{equation}
The operator $\mathbf{K}$ is defined such that $\mathbf{K}(\mathbf{H}_{0}-\epsilon_{0})=1$
everywhere except for the subspace spanned by the harmonic ground
state wave function $\phi_{0}$, where the inverse of $\mathbf{H}_{0}-\epsilon_{0}$
would be undefined and it is convenient to choose $\mathbf{K=}0$.
Contracting $\mathbf{K}$ with eq.~(\ref{eq:perturbOrdPTensor}) gives\begin{equation}
\mathbf{a}_{p}=\mathbf{K}\sum_{i=1}^{p}(\epsilon_i-\mathbf{H}_i)\mathbf{a}_{p-i}.
\label{eq:wfRecursiveRel}\end{equation}
Together equations (\ref{eq:energyRecursiveRel}) and (\ref{eq:wfRecursiveRel})
allow us to compute the ground state energy and wave function to any
order.%

We implemented the steps required to compute the $1/D$ expansion
to arbitrary order in {\em Mathematica} \cite{mathematica} and
in C++. The determination of the Taylor expansion coefficients of
the Hamiltonian is done with {\em Mathematica}. The computation of the perturbation
series (eqs.~(\ref{eq:energyRecursiveRel}) and (\ref{eq:wfRecursiveRel}))
is done in C++, for its speed of 
operations with large arrays (corresponding to the various tensor
contractions in these equations). We have computed $20$ $1/D$ expansion
coefficients (which required expanding up to order $41$ in perturbation
theory). The result is presented in Table \ref{tab:Energy-expansion-coeffients-latest}.
Also in this table are the corresponding coefficients for helium (from
an identical calculation), which agree to at least
five significant figures with Table I of \cite{goodson:8481} (after
accounting for a difference in units, which amounts to diving by $Z^{2}=4$)
and serve as a check of our calculations. Note that the coefficients in 
Table \ref{tab:Energy-expansion-coeffients-latest} become large at
high orders. This is due to the essential singularity at $\delta=0$.
The nature of this singularity has been investigated in Ref. \cite{goodson:8481}. 

\begin{figure}[htb]
\begin{centering}
\if\withFigures1
\includegraphics{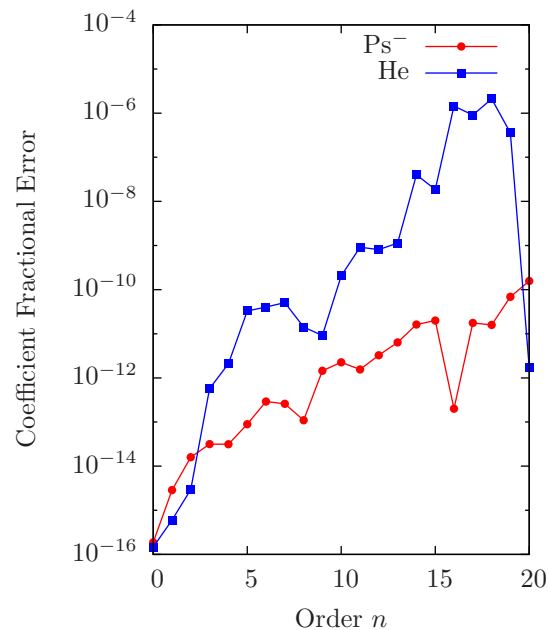}
\fi
\par\end{centering}

\caption{(Color online) Fractional error in the $1/D$ expansion coefficients defined as $|(E_{n}^{(16)}-E_{n}^{(20)})/E_{n}^{(20)}|$
as a function of the order $n$. The coefficients $E_{n}^{(16)}$
were obtained using the standard double precision arithmetic ($\approx16$
digits of precision), while $E_{n}^{(20)}$ were obtained using the
ARPREC arbitrary precision library (with $20$ digits of precision).
\label{fig:accuracy}}

\end{figure}

Due to the large number
of algebraic operations required to generate the $1/D$ expansion
we need to check for round off error in our coefficients; one way
to do this is to repeat the C++ computation at higher precision (there
should be no need to redo the {\em Mathematica} part, since {\em Mathematica}
does arbitrary precision computations by default, as long as one does
not invoke numerical solvers). We have implemented a version of the
C++ code using the arbitrary precision arithmetic package ARPREC \cite{Bailey02arprec:an}.%
\begin{table}
\begin{ruledtabular}
\begin{tabular}{crr}
%\hline 
$p$ & $\epsilon_{p}$ for $\mbox{Ps}{}^{-}$ & $\epsilon_{p}$ for $\mbox{He}$\tabularnewline
%\hline
%\hline 
$0$ & $-1.185438078904337(1)$ E0 & $-2.423036748379509(1)$ E1\tabularnewline
%\hline 
$2$ & $-2.78770519314798(1)$ E0 & $-3.544873487874171(2)$ E1\tabularnewline
%\hline 
$4$ & $-7.2695509791874(1)$ E0 & $-5.56025516084019(2)$ E1\tabularnewline
%\hline 
$6$ & $6.7347904088005(2)$ E1 & $-2.174685942637(1)$ E1\tabularnewline
%\hline 
$8$ & $-2.1412953632562(1)$ E3 & $-3.30958097736(1)$ E2\tabularnewline
%\hline 
$10$ & $7.8884951280128(7)$ E4 & $5.2508188805(2)$ E2\tabularnewline
%\hline 
$12$ & $-3.519438146299(1)$ E6 & $4.0504015254(2)$ E4\tabularnewline
%\hline 
$14$ & $1.842029153744(1)$ E8 & $-1.7333557830(1)$ E6\tabularnewline
%\hline 
$16$ & $-1.107117203726(1)$ E10 & $5.6857880174(1)$ E7\tabularnewline
%\hline 
$18$ & $7.51651405108(1)$ E11 & $-1.77525788344(2)$ E9\tabularnewline
%\hline 
$20$ & $-5.69017274005(1)$ E13 & $5.528541546(1)$ E10\tabularnewline
%\hline 
$22$ & $4.75290730575(1)$ E15 & $-1.732045588(2)$ E12\tabularnewline
%\hline 
$24$ & $-4.34262756760(1)$ E17 & $5.409411228(4)$ E13\tabularnewline
%\hline 
$26$ & $4.30867078063(3)$ E19 & $-1.638399579(2)$ E15\tabularnewline
%\hline 
$28$ & $-4.6135902290(1)$ E21 & $4.4914966(2)$ E16\tabularnewline
%\hline 
$30$ & $5.3029755999(1)$ E23 & $-8.6523653(2)$ E17\tabularnewline
%\hline 
$32$ & $-6.512769125044(1)$ E25 & $-1.289717(2)$ E19\tabularnewline
%\hline 
$34$ & $8.5114570184(2)$ E27 & $3.152243(3)$ E21\tabularnewline
%\hline 
$36$ & $-1.17941098599(2)$ E30 & $-2.796124(6)$ E23\tabularnewline
%\hline 
$38$ & $1.7272349225(1)$ E32 & $2.031490(1)$ E25\tabularnewline
%\hline 
$40$ & $-3.253630209(1)$ E34 & $-1.705787919969(3)$ E33\tabularnewline
%\hline
\end{tabular}
\par\end{ruledtabular}

\caption{$1/D$ energy expansion coefficients in eq.~(\ref{eq:energyExpansion}),
 in units of $m\alpha^{2}$. Terms with odd $p$ vanish. 
%Numbers following letter E are powers of $10$ multiplying the entries. 
Letter E indicates powers of 10 multiplying the entries. The uncertainty
in each coefficient was estimated using a 
similar calculation with $20$ digit precision as described in the text. 
\label{tab:Energy-expansion-coeffients-latest}}

\end{table}
The relative effect of round off error is shown in Fig.~\ref{fig:accuracy}.
We see that the error introduced by finite precision arithmetic is much 
smaller than the accuracy of the final ground state energy obtained by
resumming the $1/D$ expansion. Higher order calculations will require
better precision when the fractional error becomes of the same order
as the accuracy required.

%\bibliographystyle{/Users/czar/Library/texmf/tex/latex/bibtexStyles/ac}
%\bibliography{/Users/czar/Documents/pro/Tables/Archive/phd}

\begin{thebibliography}{10}

\bibitem{PhysRevA.24.3242}
A.~P. Mills,
\newblock Phys. Rev. A {\bf 24}, 3242 (1981).

\bibitem{PhysRevA.80.054502}
J.-M. Richard,
\newblock Phys. Rev. A {\bf 80}, 054502 (2009).

\bibitem{Basu2011}
A.~Basu,
\newblock in press in  Eur.~Phys.~J.~D, 
\newblock http://dx.doi.org/10.1140/epjd/e2011-20277-x (2011).

\bibitem{Kar2011119}
S.~Kar and Y.~Ho,
\newblock Comp.~Phys.~Comm. {\bf 182}, 119  (2011).

\bibitem{wheeler1946}
J.~A. Wheeler,
\newblock Ann. N. Y. Acad. Sci. {\bf 48}, 219 (1946).

\bibitem{mills1981}
A.~P. {Mills, Jr.},
\newblock Phys. Rev. Lett. {\bf 46}, 717 (1981).

\bibitem{bhatia1983}
A.~K. Bhatia and R.~J. Drachman,
\newblock Phys. Rev. A {\bf 28}, 2523 (1983).

\bibitem{PhysRevA.48.4780}
Y.~K. Ho,
\newblock Phys. Rev. A {\bf 48}, 4780 (1993).

\bibitem{Frolov99}
A.~M. Frolov,
\newblock Phys. Rev. A {\bf 60}, 2834 (1999).

\bibitem{frolov2006}
A.~M. Frolov,
\newblock Phys. Rev. E {\bf 74}, 027702 (2006).

\bibitem{Frolov2007}
A.~M.~Frolov, J.~Phys.~A {\bf 40}, 6175 (2007).

\bibitem{drake}
G.~W.~F. Drake and M.~Grigorescu,
\newblock J.~Phys.~B {\bf 38}, 3377 (2005).

\bibitem{Puchalski:2007ck}
M.~Puchalski, A.~Czarnecki, and S.~G. Karshenboim,
\newblock Phys. Rev. Lett. {\bf 99}, 203401 (2007).

\bibitem{PhysRevA.75.062510}
A.~K. Bhatia and R.~J. Drachman,
\newblock Phys. Rev. A {\bf 75}, 062510 (2007).

\bibitem{canexp}
F.~Fleischer {\em et~al.},
\newblock Can.~J.~Phys. {\bf 83}, 413 (2005).

\bibitem{mills1983}
A.~P. {Mills, Jr.},
\newblock Phys. Rev. Lett. {\bf 50}, 671 (1983).

\bibitem{fleischer2006}
F.~Fleischer {\em et~al.},
\newblock Phys. Rev. Lett. {\bf 96}, 063401 (2006).

\bibitem{FleischerInKarsh}
F.~Fleischer,
\newblock The negative ion of positronium: Decay rate measurements and
  prospects for future experiments,
\newblock in  S.~Karshenboim (ed.), {\em Precision Physics of Simple Atoms and Molecules}, 
 Lecture Notes in Physics Vol. 745, pp. 261--281, Springer,
  Berlin, 2008.

\bibitem{dimscale}
D.~R. Herschbach, J.~Avery, and O.~Goscinski (eds.),
\newblock {\em Dimensional Scaling in Chemical Physics} (Kluwer Academic
  Publishers, Dordrecht, 1992).

\bibitem{dunn:5987}
M.~Dunn {\em et~al.},
\newblock J. Chem. Phys. {\bf 101}, 5987 (1994).

\bibitem{goodson:8481}
D.~Z. Goodson, M.~L\'{o}pez-Cabrera, D.~R. Herschbach, and J.~D. {Morgan III},
\newblock J.~Chem.~Phys. {\bf 97}, 8481 (1992).

\bibitem{PhysRevA.46.5428}
D.~Z. Goodson and D.~R. Herschbach,
\newblock Phys. Rev. A {\bf 46}, 5428 (1992).

\bibitem{PhysRevA.51.R5}
D.~K. Watson and D.~Z. Goodson,
\newblock Phys. Rev. A {\bf 51}, R5 (1995).

\bibitem{witten1980}
E.~Witten,
\newblock Physics Today {\bf 33}, 38 (1980).

\bibitem{PhysRevA.34.2654}
D.~J. Doren and D.~R. Herschbach,
\newblock Phys. Rev. A {\bf 34}, 2654 (1986).

\bibitem{McGuire1964}
J.~B. McGuire,
\newblock J.~Math.~Phys. {\bf 5}, 622 (1964).

\bibitem{Lieb1963}
E.~H. Lieb and W.~Liniger,
\newblock Phys. Rev. {\bf 130}, 1605 (1963).

\bibitem{Yang1967}
C.~N. Yang,
\newblock Phys. Rev. Lett. {\bf 19}, 1312 (1967).

\bibitem{Craig1992}
T.~W. Craig, D.~Kiang, and A.~Ni\'egawa,
\newblock Phys. Rev. A {\bf 46}, 2271 (1992).

\bibitem{Li1995}
Y.-Q. Li and Z.-S. Ma,
\newblock Phys. Rev. B {\bf 52}, R13071 (1995).

\bibitem{rosenthal}
C.~M. Rosenthal,
\newblock J. Chem. Phys. {\bf 55}, 2474 (1971).

\bibitem{twoelec1d}
D.~J. Doren and D.~R. Herschbach,
\newblock J. Chem. Phys. {\bf 87}, 433 (1987).

\bibitem{Hogaasen:2010}
H.~{H{\o}gaasen}, J.-M. Richard, and P.~Sorba,
\newblock Am.~J.~Phys. {\bf 78}, 86 (2010).

\bibitem{PhysRevA.71.052505}
T.~Li and R.~Shakeshaft,
\newblock Phys. Rev. A {\bf 71}, 052505 (2005).

\bibitem{elout:5112}
M.~O. Elout {\em et~al.},
\newblock J.~Math.~Phys. {\bf 39}, 5112 (1998).

\bibitem{PadeEncyclopedia}
{G. A. Baker, Jr.} and P.~Graves-Morris,
\newblock {\em Pad\'e approximants}, 2nd ed. (Cambridge Univ.~Press, Cambridge,
  UK, 1996).

\bibitem{korobov2002}
V.~I. Korobov,
\newblock Phys. Rev. A {\bf 66}, 024501 (2002).

\bibitem{mathematica}
S.~Wolfram,
\newblock {\em Mathematica 7} ({Wolfram Research Inc.}, Champaign, Illinois,
  2008).

\bibitem{Bailey02arprec:an}
D.~H. Bailey, Y.~Hida, X.~S. Li, and O.~Thompson,
\newblock {ARPREC}: An arbitrary precision computation package,
\newblock http://crd.lbl.gov/~dhbailey/dhbpapers/arprec.pdf, 2002.



\end{thebibliography}

\end{document}